\documentclass[twocolumn]{aastex63}
\usepackage{graphics,epsf}
\usepackage{amsmath}                
\usepackage{amsfonts}               
\usepackage{amssymb}                
\usepackage{epsfig}                 
\usepackage{graphicx}
\usepackage{float}
\usepackage{color}
\usepackage[para,online,flushleft]{threeparttable}

\newcommand{\km}{{~\rm km}}
\newcommand{\s}{{~\rm s}}

\newcommand{\g}{{~\rm g}}
\newcommand{\G}{{~\rm G}}
\newcommand{\K}{{~\rm K}}
\newcommand{\erg}{{~\rm erg}}
\newcommand{\yr}{{~\rm yr}}

	\def \aap{A\&A}
	
	\def \apj{ApJ}
	\def \apjl{ApJ}
	\def \apjs{ApJS}
	\def \nat{Nature}
	
	\def \mnras{MNRAS}

\begin{document}

\title{The response of a red supergiant to a common envelope jets supernova (CEJSN) impostor event}

\author{Nitzan Ragoler}
\affiliation{Department of Physics, Technion, Haifa, 3200003, Israel; \\ 	
ealealbh@gmail.com; ronsr@physics.technion.ac.il, shlomi.hillel@gmail.com, soker@physics.technion.ac.il}

\author{Ealeal Bear}
\affiliation{Department of Physics, Technion, Haifa, 3200003, Israel; \\ 	
ealealbh@gmail.com; ronsr@physics.technion.ac.il, shlomi.hillel@gmail.com, soker@physics.technion.ac.il}

\author{Ron Schreier}
\affiliation{Department of Physics, Technion, Haifa, 3200003, Israel; \\ 	
ealealbh@gmail.com; ronsr@physics.technion.ac.il, shlomi.hillel@gmail.com, soker@physics.technion.ac.il}

\author{Shlomi Hillel}
\affiliation{Department of Physics, Technion, Haifa, 3200003, Israel; \\ 	
ealealbh@gmail.com; ronsr@physics.technion.ac.il, shlomi.hillel@gmail.com, soker@physics.technion.ac.il}

\author[0000-0003-0375-8987]{Noam Soker}
\affiliation{Department of Physics, Technion, Haifa, 3200003, Israel; \\ 	
ealealbh@gmail.com; ronsr@physics.technion.ac.il, shlomi.hillel@gmail.com, soker@physics.technion.ac.il}

\begin{abstract}
Using a one-dimensional stellar evolution code we simulate the response of a red supergiant (RSG) star to injection of energy and to mass removal. We take the values of the energy that we inject and the mass that we remove according to our previous three-dimensional hydrodynamical simulations of a neutron star (NS) on a highly eccentric orbit that enters the envelope of an RSG star for half a year and launches jets as it accretes mass via an accretion disk. We find that for injected energies of $\simeq 10^{47} - 10^{48} \erg$ and removed mass of $\simeq 0.03-0.6 M_\odot$ the RSG envelope expands to a large radius. Therefore, we expect the NS to continue to orbit inside this massive inflated envelope for several more months, up to about twice the initial RSG radius, to continue to accrete mass and launch jets for a prolonged period. Although these late jets are weaker than the jets that the NS launches while inside the original RSG envelope, the late jets might actually be more influential on the light curve, leading to a long, several months to few years, and bright, about $\ga 10^8 L_\odot$, transient event.  The RSG returns to more or less a relaxed structure after about ten years, and so another transient event might occur in the next periastron passage of the NS. Our results add to the already rich variety of jet-driven explosions/outbursts that might account for many puzzling  transient events.
\end{abstract}


\section{INTRODUCTION}
\label{sec:intro}

In a common envelope jets supernova (CEJSN) impostor event a neutron star (NS) or a black hole (BH) orbits inside the extended envelope of a red supergiant (RSG) star and launches jets as it accretes mass from the envelope through an accretion disk (e.g., \citealt{Gilkisetal2019, LopezCamaraetal2019, LopezCamaraetal2020MN, Gricheneretal2021, Schreieretal2021}). The impostor implies that the NS/BH does not enter or merge with the core of the RSG, unlike the case in a CEJSN event where the NS/BH destroys the core, accretes a fraction of its mass, and launches energetic jets; e.g, \citealt{SokeretalGG2019, GrichenerSoker2019a, Schroderetal2020, GrichenerSoker2021, Soker2021Triple}). 
 (In case the NS does not destroy the entire star and ends at its center the system forms a Thorne-Zytkow object; \citealt{ThorneZytkow1977}.)  
The NS/BH might perform a full common envelope evolution (CEE) with the RSG and eject the entire envelope as it spirals-in inside the envelope, or the NS/BH might enter the envelope and exit from it in case the NS/BH is on a highly eccentric orbit. 
CEJSNe and CEJSN impostors might involve also triple-star interaction (e.g., \citealt{Soker2021Triple, Soker2021NSNS, AkashiSoker2021}). 

Cooling by neutrino emission allows high accretion rates of $\dot M_{\rm acc} \ga 10^{-3} M_\odot \yr^{-1}$ \citep{HouckChevalier1991, Chevalier1993, Chevalier2012}, and the density gradient within the RSG envelope implies that the accreted mass has a net angular momentum that forms an accretion disk around the NS/BH (e.g.,  \citealt{ArmitageLivio2000, Papishetal2015, SokerGilkis2018}). Numerical simulations show that the energy that the jets deposit to the envelope results in mass removal from the envelope and envelope inflation (e.g., \citealt{Schreieretal2021, Hilleletal2022}). As well, the jets reduce the density in the vicinity of the NS/BH and by that reduce the accretion rate and therefore the jets' power (e.g., \citealt{LopezCamaraetal2019}), in what is generally termed the jet feedback mechanism (e.g., \citealt{Soker2016Rev}). The accretion rate cannot be too low because the neutrino cooling becomes inefficient. For detailed discussion of the accretion rate in relation to the Bondi-Hoyle-Lyttleton accretion rate and on quantifying the effect of the jet feedback mechanism in CEJSNe see \cite{Gricheneretal2021} and \cite{Hilleletal2022}.
The efficient ejection of envelope gas by the jets (e.g., \citealt{Shiberetal2019}) increases the CEE efficiency parameter, and it might become $\alpha_{\rm CE} > 1$. Indeed, some scenarios do require the CEE efficiency parameter to be above unity (e.g. \citealt{Fragosetal2019, Zevinetal2021, Garciaetal2021}).

The interaction of the jets with the envelope forms a hot region from the shocked jets and envelope gas, the so called `cocoon' (e.g., \citealt{LopezCamaraetal2019}). A fraction of the thermal energy of the cocoon ends in an optical outburst that might mimic a core collapse supernova or a core collapse supernova impostor (e.g., \citealt{Schreieretal2021}). For that, CEJSN and CEJSN impostor events might account for enigmatic transients. \cite{Thoneetal2011} propose the merger of a NS with a helium star for the unusual gamma-ray burst GRB~101225A.  \cite{SokerGilkis2018} proposed a CEJSN event to explain iPTF14hls-like transient events \citep{Arcavietal2017}, including SN~2020faa \citep{Yangetal2020}. 
\cite{SokeretalGG2019} and \cite{Soker2022FBOTs} propose that CEJSNe and CEJSN impostors, respectively, might account for some fast blue optical transients. 
\cite{Schroderetal2020} propose that SN1979c and SN1998s were CEJSN events. 
\cite{Dongetal2021} adopted processes and ingredients from the early CEJSN studies to propose the CEJSN scenario for the luminous radio transient VT~J121001+495647. 
As well, some high-energy processes might take place in the jets of CEJSN events, including  r-process nucleosynthesis in the case of a NS that spirals-in inside the RSG core \citep{GrichenerSoker2019a, GrichenerSoker2019b, Gricheneretal2022} and the formation of very high-energy neutrinos by BHs that launch jets in the deep envelope of RSG stars \citep{GrichenerSoker2021}.  
 
In this study we simulate the evolution of a RSG star after a NS on a highly eccentric orbit has passes through the envelope and deposited energy to the envelope and removed some envelope mass by the jet it launched. We take the amount of energy that the NS deposited to the envelope and the mass it removed from the envelope according to the three-dimensional hydrodynamical simulations of \cite{Schreieretal2021}. We use these quantities in a one-dimensional stellar evolutionary code (section \ref{sec:Numerical}) to follow the evolution of the star for many years. We present the results in section \ref{sec:Results}. Our summary and discussion are section \ref{sec:Summary}.

\section{Numerical set up}
\label{sec:Numerical}

To follow the evolution of the RSG star after the passage of the NS through its envelope we conduct simulations with the spherically symmetric stellar evolution code \textsc{mesa-star} (version 10398;
 \citealt{Paxtonetal2011,Paxtonetal2013,Paxtonetal2015,Paxtonetal2018,Paxtonetal2019}).

We follow the inlist of \cite{Gricheneretal2021} and divide the run to 3 parts as follows. 
In part A we follow a zero age main sequence (ZAMS) star of mass $M_{2\rm ,ZAMS}=15M_\odot$ to the RSG phase when it reaches a radius of $R_{\rm RSG} = 795 R_\odot$ and its mass is $M_2=13.2 M_\odot$.

In part B we inject energy into the RSG envelope and remove mass (for simulations of energy injection into the envelope of RSGs but in pre-explosion RSGs see, e.g., \citealt{McleySoker2014, Koetal2022}). We take the values of energy we inject and the mass that we remove from the envelope from the three-dimensional hydrodynamical simulations of \cite{Schreieretal2021}. We conduct simulations with three pairs of values of the injected energy and removed mass, $E_{\rm inj}=1.82\times 10^{46} \erg$ with $M_{\rm rem}=0.0014M_\odot$, $E_{\rm inj}=1.82\times 10^{47} \erg$ with $M_{\rm rem}=0.03M_\odot$, and $E_{\rm inj}=1.82\times 10^{48} \erg$ with $M_{\rm rem}=0.59M_\odot$.
We inject the energy at a constant (in time) power into the envelope zone of $350R_\odot < r < R_2(t)$ and remove the mass at a constant rate for a duration of either $t_{\rm inj}=1 \yr$ or $t_{\rm inj}=2 \yr$. We inject the energy at each time step with an equal energy per unit mass in the above envelope zone. Note that $R_2(t)$ increases as we inject the energy. Because of numerical difficulties we could not simulate higher values of injected energy. Specifically, when we inject an energy of $E_{\rm inj}\ga 3\times 10^{48} \erg$ the stellar model does not converge. 

We name each simulation by the logarithm of the injected energy in $\erg$, the mass removed in $M_\odot$, and the injection time in years. For example, S(48,0.59,2) is the simulations with $E_{\rm inj}=1.82\times 10^{48} \erg$, $M_{\rm rem}=0.59 M_\odot$, and $t_{\rm inj}=2 \yr$.

In part C we simulate the evolution from the time energy deposition ceased up to $t=16 \yr$ (the orbital period is $T=16.6\yr$). We follow the same inlist as in part A , where no energy is deposited.

We take the energy injection and mass removal time $t_{\rm inj}$ to be about one to two years for two reasons. 
The first one is physical. Due to the highly eccentric orbit, the jets that the NS launches deposit the energy into the envelope only on the side of the RSG star along the NS orbit inside the envelope. The NS spends about half a year inside the envelope. It would take convection to carry the energy to the entire star a time that is somewhat longer than the dynamical time of the RSG star $\tau_{\rm d} = 0.66 \yr$. Namely, within about one-few years. We are using a spherical code and should therefore consider this time to spread the energy. The second reason is numerical. The numerical code cannot converge on a stellar model for shorter time scales of energy injection for the values of $E_{\rm inj}$ that we simulate here. 

\section{Results}
\label{sec:Results}

\subsection{Envelope inflation}
\label{subsec:EnvelopeInflation}
We present the evolution with time of the RSG radius, its density profile, and its mass profile as a result of the energy injection and mass removal that we described in section \ref{sec:Numerical}. We recall that we present cases with injected energies and removed mass of $E_{\rm inj}=1.82\times 10^{46} \erg$ with $M_{\rm rem}=0.0014M_\odot$, of $1.82\times 10^{47} \erg$ with $M_{\rm rem}=0.03M_\odot$, and of $1.82\times 10^{48} \erg$ with $M_{\rm rem}=0.59M_\odot$. These values are according to the numerical results of \cite{Schreieretal2021} who simulated the passage of a NS on a highly eccentric orbit inside the envelope of a RSG star. The NS reaches a periastron distance of $r_{\rm p}=400 R_\odot$ and spends half a year inside the volume of the unperturbed RSG envelope. \cite{Schreieretal2021} simulated cases with more energetic jets, but we could not reach convergence with our numerical tools for these very energetic cases. 

In the first two cases above, the energy deposition and mass removal last for either $t_{\rm inj}=1 \yr$ or $t_{\rm inj}=2 \yr$, while for the most energetic case we simulate we reach no convergence for $t_{\rm inj}=1 \yr$, only for $t_{\rm inj}=2 \yr$.  

\begin{figure}[ht]
\includegraphics[trim= 3.3cm 9.4cm 0.0cm 9.4cm, scale=0.64]{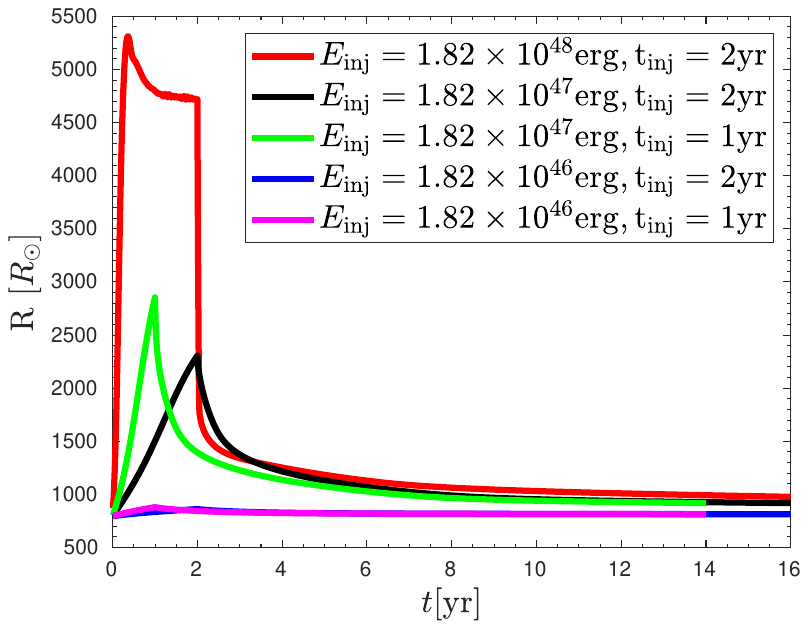}
\caption{Radius versus time for $E_{\rm inj}=1.82 \times 10^{46} \erg, 1.82 \times 10^{47} \erg, 1.82 \times 10^{48} \erg$ and mass loss of $M_{\rm lost}=0.0014M_\odot, 0.03M_\odot, 0.59M_\odot$ respectively. The energy deposition and the mass removal both occur simultaneously for a period of $t_{\rm inj}$ starting at $t=0$ (when the NS enters the RSG envelope), as written in the legend. The red, black, green, blue, and purple lines are for simulations S(48,0.59,2), S(47,0.03,2), S(47,0.03,1), S(46,0.0014,2), and S(46,0.0014,1), respectively. 
 }
 \label{Fig:r_vs_t_combined}
 \end{figure}

In Fig. \ref{Fig:r_vs_t_combined} we present the RSG radius as a function of time for these five cases. We learn that at the moment we stop the energy injection the RSG radius decreases very rapidly, but not completely to its original value of $R_2=795 R_\odot$. 

For the two cases with $E_{\rm inj}=1.82\times 10^{46} \erg$ the RSG structure does not change much, and the RSG reaches a maximum radius of $R_{\rm 2,max}=880 R_\odot$ for S(46,0.0014,1) and $R_{\rm 2,max}=860 R_\odot$ for S(46,0.0014,1). We will not present the density and mass profiles for these cases as they are of less interest to us. 

\begin{figure}[ht]
\includegraphics[trim= 3.3cm 9.4cm 0.0cm 9.4cm, scale=0.64]{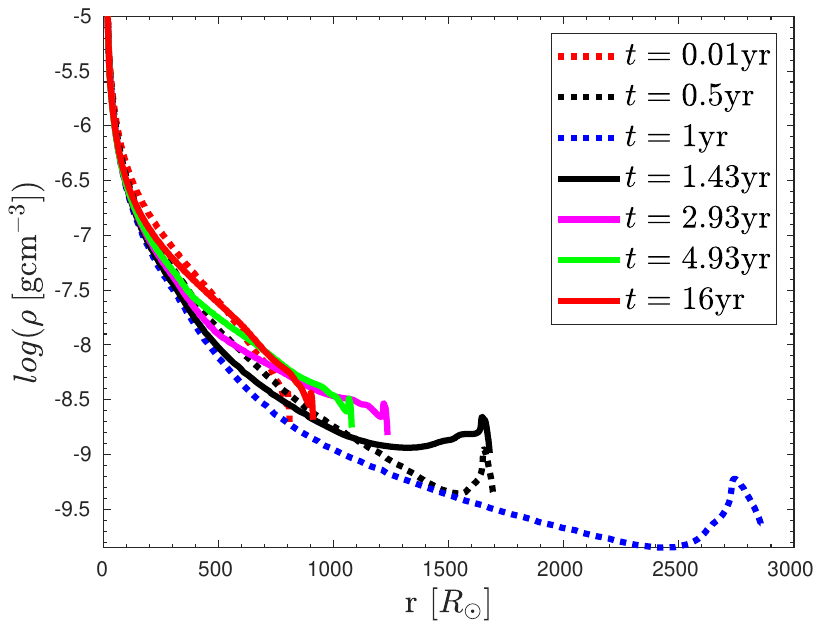}
\caption{Density versus radius for simulation S(47,0.03,1), i.e., with injected energy of $E_{\rm inj}= 1.82 \times 10^{47} \erg $ and mass removal of $M_{\rm lost}=0.03M_\odot$. The energy deposition and the mass removal both occur simultaneously for a period of $t_{\rm inj}=1\yr$ starting at $t=0$. The dotted lines are for the times when energy is deposited. The solid lines are for times after energy is deposited as written in the legend. }
 \label{Fig:Rho_vs_r_d47}
 \end{figure}

\begin{figure}[ht]
\includegraphics[trim= 3.4cm 9.4cm 0.0cm 9.4cm, scale=0.64]{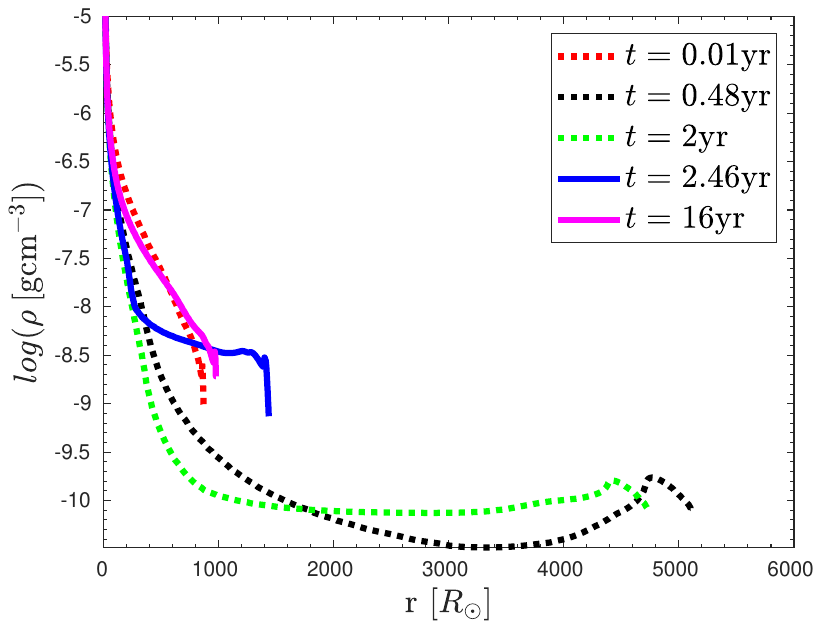}
\caption{Similar to Fig. \ref{Fig:Rho_vs_r_d47} but for simulation S(48,0.59,2) and at different times.  }
 \label{Fig:Rho_vs_r_d48}
 \end{figure}

\begin{figure}[ht]
\includegraphics[trim= 3.4cm 9.4cm 0.0cm 9.5cm, scale=0.64]{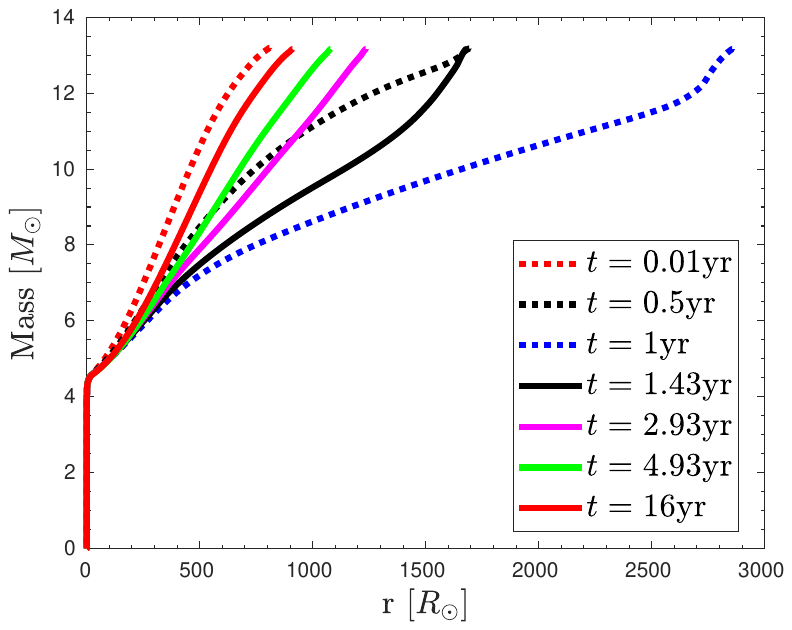}
\caption{Mass $M(r)$ inner to radius $r$ as function of $r$ for simulation S(47,0.03,1) that we present in Fig. \ref{Fig:Rho_vs_r_d47} and at the same times. Dotted lines are for times during the energy injection and mass removal phase.  }
 \label{Fig:m_vs_r_d47}
 \end{figure}

\begin{figure} 
\includegraphics[trim= 3.4cm 9.4cm 0.0cm 9.5cm, scale=0.64]{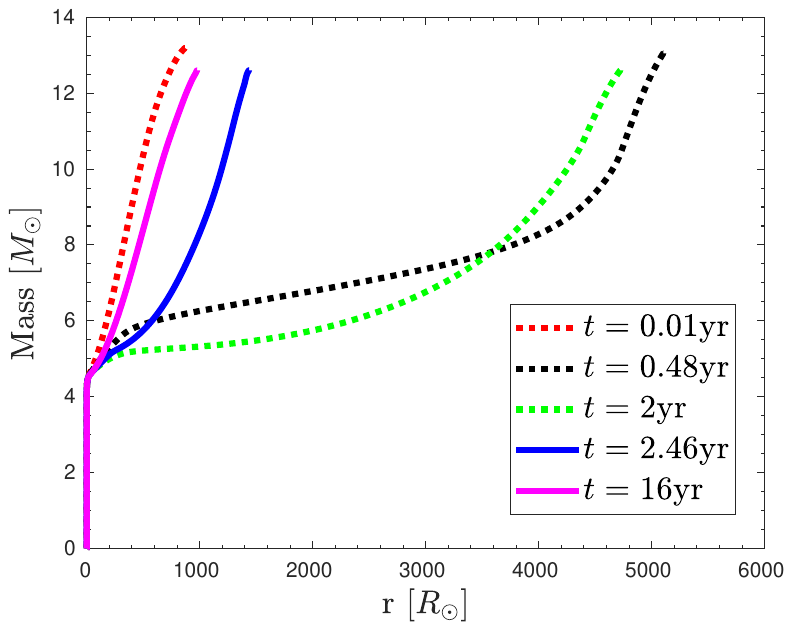}
\caption{Similar to Fig. \ref{Fig:m_vs_r_d47} but for simulation S(48,0.59,2) and at different times. 
 }
 \label{Fig:m_vs_r_d48}
 \end{figure}

In Fig. \ref{Fig:Rho_vs_r_d47} and \ref{Fig:Rho_vs_r_d48} we present the density profiles at several times for simulations S(47,0.03,1) and S(48,0.59,2), respectively. In Figs. \ref{Fig:m_vs_r_d47} and \ref {Fig:m_vs_r_d48} we present the mass $m(r)$ inner to radius $r$ as function of $r$ for several times of simulations S(47,0.03,1) and S(48,0.59,2), respectively. Because an energy injection phase of $t_{\rm inj}=2 \yr$ is much longer than the passage of the NS inside the RSG envelope, we will analyse in detail only the results of simulation S(47,0.03,1). 

\subsection{On the assumptions of the 1D models}
\label{subsec:Assumptions}

Conducting the simulations with \textsc{mesa} implies two strong assumptions, that the star is spherically symmetric and that the stellar model is in hydrostatic equilibrium. Simply, the numerical code \textsc{mesa} (section \ref{sec:Numerical}) assumes a hydrostatic equilibrium of a spherically symmetric star.
Neither of these two are correct for the present highly non-spherical flow that result from the strong jets that the NS launches into the envelope \citep{Schreieretal2021}.

Strictly speaking, our results are not consistent when the expansion timescale of the star is shorter than the dynamical timescale at the given radius. This is the case during the first $\simeq 1.5 \yr$ when the stellar radius is larger than $\simeq 1500 R_\odot$ in simulation S(47,0.03,1) and for the first $\simeq 2.5 \yr$ in simulation S(48,0.59,2).  The dynamical time of the RSG is $\tau_{\rm d} = (G \bar \rho)^{-1/2} \simeq 0.9 (R_2/1000 R_\odot)^{3/2} \yr$, where $\bar \rho$ is the average density of the RSG (including the core). We can therefore trust the results of simulation S(47,0.03,1) from $t \simeq 1.4 \yr$ when the radius has contracted back to $R_2 \simeq 1700$ (see solid-black line in Figs. \ref{Fig:Rho_vs_r_d47} and \ref{Fig:m_vs_r_d47} at $t=1.43 \yr$). For simulation S(48,0.59,2) we see from Figs. 
\ref{Fig:Rho_vs_r_d48} and \ref{Fig:m_vs_r_d48} that about half a year after the end of the energy injection  and mass removal (i.e., at $t\simeq 2.5 \yr$) the radius rapidly decreases to $R_2=1400 R_\odot$. 

\begin{figure} 
\includegraphics[trim= 3.9cm 8.5cm 0.0cm 8.5cm, scale=0.64]{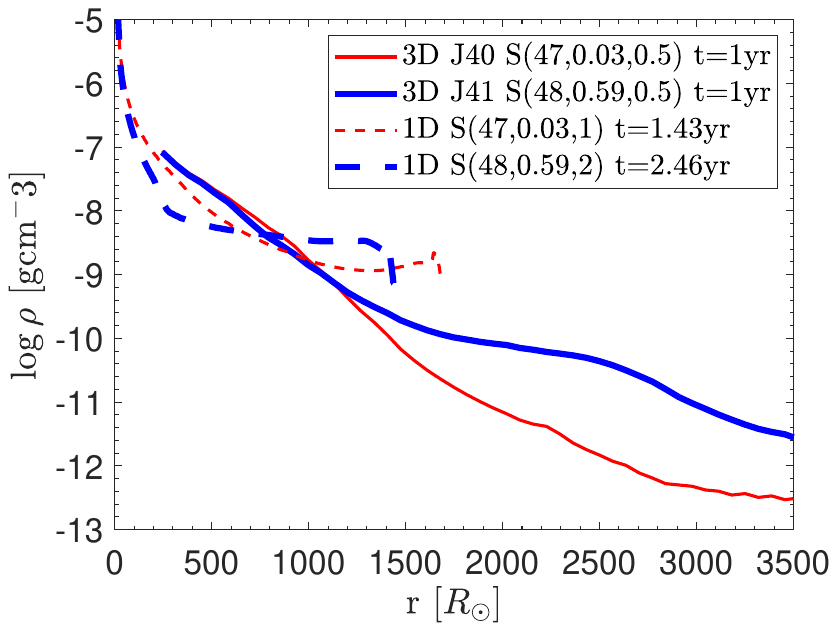}
\caption{ {{{ Comparing the density profiles of the present 1D simulations (dotted lines) with the average density of the 3D hydrodynamical simulations (solid lines) of \cite{Schreieretal2021}. All plots are at about half a year after the end of energy injection (see inset for the specific times). The thin-red lines are for the S(47,0.03,1) 1D model and its equivalent 3D model, and the thick-blue lines are for the S(48,0.59,2) 1D model and its equivalent 3D model. }}} 
 }
 \label{Comparison1D3D}
 \end{figure}

 To show the qualitative agreement between the 1D and 3D simulations regarding the inflated envelope, we present in Fig. \ref{Comparison1D3D} the density profiles from the 1D simulations of models S(47,0.03,1) and S(48,0.59,2), with their respective equivalents from the 3D simulations of \cite{Schreieretal2021}. All lines are at about half a year after the end of energy injection. For the 3D simulations this is at about $t=1 \yr$, for S(47,0.03,1) it is at about $t=1.5 \yr$, and for S(48,0.59,2) it is at about $t=2.5 \yr$ (see inset of Fig. \ref{Comparison1D3D} for the exact times).   

Despite these limitations we do find that our simulations have merit and present new results. 

Firstly, our 1D simulations validate the finding from the 3D simulations \citep{Schreieretal2021} of an extended envelope to a  radius of about twice the initial RSG radius, $R_2 \simeq 2 R_{\rm RSG}$. The advantage of our simulations is that we do so with a full stellar code (for example, the 3D simulations did not include radiative transfer and convection). Since at these times the stellar zones $r \la 2 R_{\rm RSG}$ had time to dynamically relax, we trust more the density profiles of our 1D simulations than those of the 3D simulations in these regions. The real densities at $r \simeq 2 R_{\rm RSG}$ might be in between the profiles of the 1D and 3D simulations.  Fig. \ref{Comparison1D3D} shows that the density of the inflated envelope out to $\simeq 2 R_{\rm RSG}$ has a large enough density to prolong the accretion phase of the NS, as we discuss in section \ref{subsec:Prolongedaccretion}.  

 Secondly, our simulations can follow the star for several years, unlike the 3D hydrodynamical-only simulations. We find that the star returns to almost its initial structure, but with less mass as we removed mass. 

Let us comment on the thermal timescale. The energy that we deposit into the RSG not only drives the star out of dynamical equilibrium but also from thermal equilibrium. However, we first note that \textsc{mesa} is built to deal with stars out of thermal equilibrium, but still in hydrostatic equilibrium (e.g., section 2.3 in \citealt{Paxtonetal2015}). For example, \textsc{mesa} calculates the pre-main sequence evolution that proceeds on a thermal time scale. Secondly, within the time that we consider the density profiles, at about half a year after the end of energy injection, the star radiates away its excess internal energy. We now show this.  

At the relevant times of about half a year after the end of energy injection, $t=1.5 \yr$ for model S(47,0.03,1) and $t=2.5 \yr$ for model S(48,0.59,2), the required stellar luminosity to make the thermal timescale about equal to the simulation time is 
\small
\begin{eqnarray}
\begin{aligned} 
L_{\rm th} & = \frac{G M^2_2}{R_2 t  } = 
3.5 \times 10^{6} 
\left(\frac{M_2}{13M_\odot} \right)^{2}
\\ & \times
\left(\frac{R_2}{1500 R_\odot} \right)^{-1}
\left(\frac{t}{1 \yr} \right)^{-1}
L_\odot,
\end{aligned}
\label{eq:Lcrit}
\end{eqnarray}
\normalsize
where we substituted typical values at the relevant times. We find from our simulations that 
the luminosities are $L(0.5\yr)= 4.1 \times 10^5 L_\odot$, $L(1\yr)= 1.1 \times 10^6 L_\odot$, and $L(1.43\yr)= 1.9 \times 10^5 L_\odot$ for model S(47,0.03,1), and $L(1\yr)= 7.4 \times 10^6 L_\odot$, $L(2\yr)= 7.5 \times 10^6  L_\odot$, and $L(2.46\yr)= 1.0 \times 10^5 L_\odot$ for model S(48,0.59,2).

Considering the approximate nature of equation (\ref{eq:Lcrit}), we conclude that our inflated stellar model manages to arrange itself thermally by half a year after energy injection ends. This is similar to many other simulations with \textsc{mesa}, e.g., in the simulations of \cite{RenzoGotberg2021} where accretion drives the star out of thermal equilibrium but the mass-accreting star radiates away its excess internal energy.

\subsection{Prolonged accretion}
\label{subsec:Prolongedaccretion}

The time $t=0$ in our simulations corresponds to the time when the NS enters the RSG envelope. Namely, when it starts to accrete mass at a high rate and to launch energetic jets. The NS reaches periastron at $t=0.255 \yr$. In their three-dimensional hydrodynamical simulations \cite{Schreieretal2021} assume that the jets accretion phase ends when the NS exits the original photosphere of the RSG, and launches jets for a time of $0.51 \yr$. However, the density distribution in their simulations show that the jets accelerate gas outwards, such that the NS star continues to move inside a relatively dense gas for a much longer time. Their simulations, however, did not include the correct equation of stellar evolution. Our results here show that as the NS continues to move away from periastron and outside the original RSG envelope, it can accrete mass from a relatively dense envelope out to $r \simeq 1500 - 1700  R_\odot$, about twice the original RSG radius. The trajectory of the NS brings it to $r=1700 R_\odot$ at $t=0.89 \yr$, i.e., about $4.5$ months after it left the original zone of the RSG envelope.

From the density profiles that we present in Fig. \ref{Fig:Rho_vs_r_d47} we learn that the density at $r=1500- 1700 R_\odot$ at $t \simeq 1.4 \yr$ is about $\rho_{1500} \simeq 0.03 \rho_{\rm p}$, where $\rho_{\rm p}$ is the initial RSG density at $r_{\rm p}=400 R_\odot$, the periastron distance of the NS.
The velocity of the NS at periastron is $v_{\rm p} = 114 \km \s^{-1}$, while it is about half this value at $r \simeq 1500-1700 R_\odot$, $v_{1500} \simeq 0.5 v_{\rm p}$. 
The Bondi-Hoyle-Lyttleton accretion rate by the NS goes as $\dot M_{\rm acc} \propto \rho v^{-3}$. From these values we find that the Bondi-Hoyle-Lyttleton accretion rate by the NS after it moves out from the initial location of the RSG photosphere and at about twice the original radius of the RSG is not much below that at periastron. Namely, $\dot M_{\rm acc,BHL} (1500 R_\odot) \approx 0.3 \dot M_{\rm acc,BHL}(r_p) \simeq 0.05 M_\odot \yr^{-1}$ for simulation S(47,0.03,1).
Note that the actual accretion rate will be $\dot M_{\rm acc} \simeq (0.01-0.1) \dot M_{\rm acc,BHL}$ due to the negative jet feedback mechanism (e.g., \citealt{Gricheneretal2021, Hilleletal2022}).

\section{Discussion and Summary}
\label{sec:Summary}

Using the one-dimensional stellar evolution code \textsc{mesa} (section \ref{sec:Numerical}) we simulated the response of RSG stellar models to injection of energy to the outer envelope and to mass removal. We list the values of the five cases in Fig. \ref{Fig:r_vs_t_combined}. We take these values from the three-dimensional hydrodynamical simulations of \cite{Schreieretal2021}. They simulated CEJSN impostor events where a NS on a highly eccentric orbit enters the RSG envelope for about half a year and then exits, to return after an orbital period of $16.6 \yr$. Because of numerical limitations we could not simulate the most energetic cases that \cite{Schreieretal2021} simulated. We therefore analysed in detail the simulation S(47,0.03,1) for which we present the density profiles and mass profiles for in Figs. \ref{Fig:Rho_vs_r_d47} and \ref{Fig:m_vs_r_d47}, respectively. 

From Fig. \ref{Fig:r_vs_t_combined}, \ref{Fig:Rho_vs_r_d47} and \ref{Fig:Rho_vs_r_d48} we see that the next time the NS enters the RSG envelope, about 16 years later, the RSG has relaxed to have more or less its original structure. The RGB has a little less mass, as we removed mass, and its radius is a little larger. But the CEJSN impostor event can repeat itself to have similar properties to the previous event. 
  
Because of the assumption of hydrostatic equilibrium in the numerical procedure we cannot trust the stellar structure when the envelope is highly inflated at early times (section \ref{subsec:EnvelopeInflation}). We can trust the structure at later times after the envelope has contracted, like at $t=1.43 \yr$ when the RSG radius is about twice its initial radius (solid-black lines in Figs. \ref{Fig:Rho_vs_r_d47} and \ref{Fig:m_vs_r_d47}). 

Our main conclusion is that because the jets that the NS launches while inside the dense RSG envelope inflate an envelope with a non-negligible mass (Fig. \ref{Fig:m_vs_r_d47} and \ref{Fig:m_vs_r_d48}), the NS continues to accrete at a relatively high rate to distances that are about twice the original RSG radius of $795 R_\odot$ (section \ref{subsec:Prolongedaccretion}). This means several more months of jet-activity. The accretion rate at these distances, of up to $r \simeq 1500-1700 R_\odot$ for simulation R(47,0.03,1), can be few tenths of the accretion rate at periastron passage of radius $r_{\rm p} =400 R_\odot$, i.e., $\dot M_{\rm acc} (1500 R_\odot) \approx 0.3 \dot M_{\rm acc}(r_p) \simeq 0.05 M_\odot \yr^{-1}$.

Despite the fact that the jets that the NS launches while orbiting in the inflated envelope are much weaker than the jets that the NS launches inside the original RSG envelope zone, these late jets might actually be more influential on the light curve. 
\cite{Schreieretal2021} estimate that the efficiency of converting the jets’ energy to radiation increases from $0.3\%$ in the lowest energy simulation they perform with $E_{\rm inj} =1.82 \times 10^{46} \erg$, to $0.7\%$ in their simulation with $E_{\rm inj} =1.82 \times 10^{47} \erg$, and to $4.3\%$ in the simulation with $E_{\rm inj} =1.82 \times 10^{49} \erg$. Most of the energy that these jets deposit to the envelope goes to perform work. As the gas expands it cools adiabatically, and therefore only a small fraction of the energy of these jets ends in radiation. The late jets that the NS launches while at larger orbital separations interact with the already ejected mass at large distances. This implies a longer expansion time and therefore the gas suffers only small adiabatic losses. In addition, the lower optical depth in the outer regions implies a shorter photon diffusion time to the photosphere, and therefore further allowing more thermal energy to be radiated away before the gas suffers adiabatic losses. \cite{KaplanSoker2020a} study the role of late jets on the light curve of core collapse supernovae and found that the efficiency of converting the energy of late jets to radiation can be up to $\approx 10-30 \%$.  

In the simulations of \cite{Schreieretal2021} the periastron distance is $r_{\rm p}=400 R_\odot$, and they simulated cases with jets' energies up to three orders of magnitude larger than in simulation S(47,0.03,1). According to the jet feedback mechanism results of \cite{Gricheneretal2021} and \cite{Hilleletal2022} we indeed expect that the jets' energy in the simulations of \cite{Schreieretal2021} be two to three orders of magnitude larger than the jets' energy in simulation S(47,0.03,1). The injected energy in simulation S(47,0.03,1) might correspond to a situation where the NS enters only the very outer region of the RSG envelope, and therefore the mass accretion rate is lower and therefore the jets' power is lower, and spends only a short time there, only few weeks. Namely, for an orbit where the periastron distance is close to the surface, i.e.,  $0<R_{2,0}-r \ll R_{2,0}$, where $R_{2,0}$ is the unperturbed radius of the RSG. Our results suggest that even in these cases the CEJSN impostor event will be bright because the NS will continue to launch jets as it orbits inside the inflated envelope for months after it has left the original zone of RSG envelope. The luminosity can be $\approx {\rm few} \times 10^{41} \erg \s^{-1} \simeq 10^8 L_\odot$ for the Bondi-Hoyle-Lyttleton accretion rate from the inflated envelope of $\dot M_{\rm acc,BHL} (1500 R_\odot) \approx 0.05 M_\odot \yr^{-1}$ that we estimated in section \ref{subsec:Prolongedaccretion}. Note that the actual accretion rate would be $(0.01-0.1)M_{\rm acc,BHL}$ because of the negative jet feedback mechanism (e.g., \citealt{Gricheneretal2021, Hilleletal2022}).   
 The peak duration might be several weeks. 

On a broader scope, our study adds to the claim that CEJSN impostors might account for puzzling-rare transient events and to the rich variety of light curves that such jet-driven events have. 

\section*{Acknowledgements}
This research was supported by the Amnon Pazy Research Foundation. We thank an anonymous referee for pointing out the limitation of the \textsc{mesa} code.

\section*{Data availability}

The data underlying this article will be shared on reasonable request to the corresponding author. 

\label{lastpage}

\begin{thebibliography}{}

\bibitem[\protect\citeauthoryear{Akashi \& Soker}{2021}]{AkashiSoker2021} Akashi M., Soker N., 2021, ApJ, 923, 55. doi:10.3847/1538-4357/ac2d2b

\bibitem[Arcavi et al.(2017)]{Arcavietal2017} Arcavi, I., Howell, D.~A., Kasen, D., et al.\ 2017, \nat, 551, 210

\bibitem[Armitage \& Livio(2000)]{ArmitageLivio2000} Armitage, P.~J., \& Livio, M.\ 2000, \apj, 532, 540

\bibitem[Baltay et al.(2013)]{Baltayetal2013}  Baltay, C., Rabinowitz, D., Hadjiyska, E., Walker, E.~S., Nugent, P., Coppi, P., Ellman, N.,  et al.\ 2013, \pasp, 125, 683. doi:10.1086/671198 
 
\bibitem[Bellm et al.(2019)]{Bellmetal2019} Bellm, E.~C., Kulkarni, S.~R., Graham, M.~J., Dekany, R., Smith, R.~M., Riddle, R., Masci, F.~J., et al.\ 2019, \pasp, 131, 018002. doi:10.1088/1538-3873/aaecbe 

\bibitem[Chamandy et al.(2018)]{Chamandyetal2018} Chamandy, L., Frank, A., Blackman, E.~G., et al.\ 2018, \mnras, 480, 1898
 
\bibitem[Chevalier(1993)]{Chevalier1993} Chevalier, R.~A.\ 1993, \apjl, 411, L33

\bibitem[Chevalier(2012)]{Chevalier2012} Chevalier, R.~A.\ 2012, \apjl, 752, L2

  
\bibitem[\protect\citeauthoryear{Dong et al.}{2021}]{Dongetal2021} Dong D.~Z., Hallinan G., Nakar E., Ho A.~Y.~Q., Hughes A.~K., Hotokezaka K., Myers S.~T., et al., 2021, Sci, 373, 1125. doi:10.1126/science.abg6037

\bibitem[Fragos et al.(2019)]{Fragosetal2019} Fragos, T., Andrews, J.~J., Ramirez-Ruiz, E., Meynet, G., Kalogera, V., Taam, R.~E., \& Zezas, A., \ 2019, \apjl, 883, L45. doi:10.3847/2041-8213/ab40d1

\bibitem[Fryer et al.(1996)]{Fryeretal1996} Fryer, C.~L., Benz, W., \& Herant, M.\ 1996, \apj, 460, 801

\bibitem[Fryxell et al.(2000)] {Fryxelletal2000} Fryxell, B., et al. 2000, \apjs, 131, 273

\bibitem[Garc{\'\i}a et al.(2021)]{Garciaetal2021} Garc{\'\i}a, F., Simaz Bunzel, A., Chaty, S., Porter, E., \& Chassande-Mottin, E.\ 2021,  \aap, 649, A114. doi:10.1051/0004-6361/202038357

\bibitem[Gilkis et al.(2019)]{Gilkisetal2019} Gilkis, A., Soker, N., \& Kashi, A.\ 2019, \mnras, 482, 4233


\bibitem[Glanz \& Perets(2021)]{GlanzPerets2021} Glanz, H. \& Perets, H.~B.\ 2021, \mnras, 507, 2659. doi:10.1093/mnras/stab2291
  
\bibitem[\protect\citeauthoryear{Grichener, Cohen, \& Soker}{2021}]{Gricheneretal2021} Grichener A., Cohen C., Soker N., 2021, ApJ, 922, 61. doi:10.3847/1538-4357/ac23dd

\bibitem[\protect\citeauthoryear{Grichener, Kobayashi, \& Soker}{2022}]{Gricheneretal2022} Grichener A., Kobayashi C., Soker N., 2022, ApJL, 926, L9. doi:10.3847/2041-8213/ac4f68

\bibitem[Grichener \& Soker(2019a)]{GrichenerSoker2019a} Grichener, A., \& Soker, N.\ 2019b, \apj, 878, 24

\bibitem[Grichener \& Soker(2019b)]{GrichenerSoker2019b} Grichener, A. \& Soker, N.\ 2019b, arXiv:1909.06328

\bibitem[Grichener \& Soker(2020)]{GrichenerSoker2020} Grichener, A. \& Soker, N.\ 2020, work presented at the EAS Annual Meeting (EWASS) 2020. 


\bibitem[Grichener \& Soker(2021)]{GrichenerSoker2021} Grichener, A. \& Soker, N.\ 2021, \mnras, 507, 1651. doi:10.1093/mnras/stab2233


\bibitem[Hillel et al.(2022)]{Hilleletal2022} Hillel, S., Schreier, R., \& Soker, N.\ 2022, \mnras, 514, 3212. doi:10.1093/mnras/stac1341

\bibitem[Holgado et al.(2021)]{Holgadoetal2021} Holgado, A.~M., Silva, H.~O., Ricker, P.~M., \& Yunes, N., \ 2021, \apjl, 910, L22. doi:10.3847/2041-8213/abecdd

\bibitem[Houck \& Chevalier(1991)]{HouckChevalier1991} Houck, J.~C., \& Chevalier, R.~A.\ 1991, \apj, 376, 234

\bibitem[Ivezi{\'c} et al.(2019)]{Ivezicetal2019} Ivezi{\'c}, {\v{Z}}., Kahn, S.~M., Tyson, J.~A., Abel, B., Acosta, E., Allsman, R., Alonso, D., et al.\ 2019, \apj, 873, 111. doi:10.3847/1538-4357/ab042c 

\bibitem[\protect\citeauthoryear{Kaplan \& Soker}{2020}]{KaplanSoker2020a} Kaplan N., Soker N., 2020, MNRAS, 492, 3013. doi:10.1093/mnras/staa020

\bibitem[\protect\citeauthoryear{Ko et al.}{2022}]{Koetal2022} Ko T., Tsuna D., Takei Y., Shigeyama T., 2022, ApJ, 930, 168. doi:10.3847/1538-4357/ac67e1

\bibitem[Kochanek et al.(2017)]{Kochaneketal2017PASP}  Kochanek, C.~S., Shappee, B.~J., Stanek, K.~Z.,  Holoien, T.~W.-S., Thompson, T.~A., Prieto, J.~L., Dong S., et al.\ 2017, \pasp, 129, 104502. doi:10.1088/1538-3873/aa80d9

\bibitem[Livio et al.(1986)]{Livioetal1986} Livio, M., Soker, N., de Kool, M., \& Savonije, G.~J., \ 1986, \mnras, 222, 235. doi:10.1093/mnras/222.2.235    
    
\bibitem[Lombardi et al.(2006)]{Lombardietal2006} Lombardi, J.~C., Jr., Proulx, Z.~F., Dooley, K.~L., Theriault, E.~M., Ivanova, N., \& Rasio, F.~A.\ 2006, \apj, 640, 441

\bibitem[L{\'o}pez-C{\'a}mara et al.(2019)]{LopezCamaraetal2019} L{\'o}pez-C{\'a}mara, D., De Colle, F., \& Moreno M{\'e}ndez, E.\ 2019, \mnras, 482, 3646

\bibitem[L{\'o}pez-C{\'a}mara et al.(2020)]{LopezCamaraetal2020MN} L{\'o}pez-C{\'a}mara, D., Moreno M{\'e}ndez, E., \& De Colle, F.\ 2020, \mnras, 497, 2057

\bibitem[MacLeod \& Ramirez-Ruiz(2015a)]{MacLeodRamirezRuiz2015a} MacLeod, M., \&  {Ramirez-Ruiz}, E.\ 2015a, \apjl, 798, L19

\bibitem[MacLeod et al.(2017)]{MacLeodetal2017} MacLeod, M., Antoni, A., {Murguia-Berthier}, A., Macias, P., \&  {Ramirez-Ruiz}, E.\ 2017, \apj, 838, 56

\bibitem[MacLeod \& Ramirez-Ruiz(2015b)]{MacLeodRamirezRuiz2015b} MacLeod, M., \& Ramirez-Ruiz, E.\ 2015b, \apj, 803, 41


\bibitem[\protect\citeauthoryear{Mcley \& Soker}{2014}]{McleySoker2014} Mcley L., Soker N., 2014, MNRAS, 445, 2492. doi:10.1093/mnras/stu1952

\bibitem[Moreno M{\'e}ndez et al.(2017)]{MorenoMendezetal2017} Moreno M{\'e}ndez, E., L{\'o}pez-C{\'a}mara, D., \& De Colle, F.\ 2017, \mnras, 470, 2929

\bibitem[Papish et al.(2015)]{Papishetal2015} Papish, O., Soker, N., \& Bukay, I.\ 2015, \mnras, 449, 288

\bibitem[Paxton et al.(2011)]{Paxtonetal2011} Paxton, B., Bildsten, L., Dotter, A., et al.\ 2011, \apjs, 192, 3

\bibitem[Paxton et al.(2013)] {Paxtonetal2013} Paxton, B., Cantiello, M., Arras, P., et al. 2013, \apjs, 208, 4

\bibitem[Paxton et al.(2015)]{Paxtonetal2015} Paxton, B., Marchant, P., Schwab, J., et al.\ 2015, \apjs, 220, 15

\bibitem[Paxton et al.(2018)]{Paxtonetal2018} Paxton, B., Schwab, J., Bauer, E.~B., et al.\ 2018, \apjs, 234, 34

\bibitem[Paxton et al.(2019)]{Paxtonetal2019} Paxton, B., Smolec, R., Schwab, J., et al.\ 2019, \apjs, 243, 10, 

\bibitem[Rasio \& Shapiro(1991)]{RasioShapiro1991} Rasio, F.~A., \& Shapiro, S.~L.\ 1991, \apj, 377, 559

\bibitem[\protect\citeauthoryear{Renzo \& G{\"o}tberg}{2021}]{RenzoGotberg2021}  Renzo M., G{\"o}tberg Y., 2021, ApJ, 923, 277. doi:10.3847/1538-4357/ac29c5 

\bibitem[Ricker \& Taam(2008)]{RickerTaam2008}  Ricker, P.~M., \& Taam, R.~E.\ 2008, \apjl, 672, L41

\bibitem[\protect\citeauthoryear{Schreier et al.}{2021}]{Schreieretal2021} Schreier R., Hillel S., Shiber S., Soker N., 2021, MNRAS, 508, 2386. doi:10.1093/mnras/stab2687

\bibitem[Schr{\o}der et al.(2020)]{Schroderetal2020} Schr{\o}der, S.~L., MacLeod, M., Loeb, A., et al.\ 2020, \apj, 892, 13

\bibitem[Shiber et al.(2019)]{Shiberetal2019} Shiber, S., Iaconi, R., De Marco, O., \& Soker, N.\ 2019, \mnras, 488, 5615. doi:10.1093/mnras/stz2013

\bibitem[Shiber et al.(2016)]{Shiberetal2016} Shiber, S., Schreier, R., \& Soker, N.\ 2016, RAA, 16, 117

\bibitem[\protect\citeauthoryear{Soker}{2016}]{Soker2016Rev} Soker N., 2016, NewAR, 75, 1. doi:10.1016/j.newar.2016.08.002

\bibitem[Soker(2021a)]{Soker2021effervescent} Soker, N.\ 2021a, \apj, 906, 1. doi:10.3847/1538-4357/abca8f

\bibitem[Soker(2021b)]{Soker2021Triple}  Soker, N.\ 2021b, \mnras. doi:10.1093/mnras/stab1275

\bibitem[Soker(2021c)]{Soker2021NSNS} Soker, N.\ 2021c, arXiv:2105.06452

\bibitem[\protect\citeauthoryear{Soker}{2022}]{Soker2022FBOTs} Soker N., 2022, in preparation.

\bibitem[Soker \& Gilkis(2018)]{SokerGilkis2018} Soker, N., \& Gilkis, A.\ 2018, \mnras, 475, 1198

\bibitem[Soker et al.(2019)]{SokeretalGG2019} Soker, N., Grichener, A., \& Gilkis, A.\ 2019, \mnras, 484, 4972

\bibitem[Th{\"o}ne et al.(2011)]{Thoneetal2011} Th{\"o}ne, C.~C., de Ugarte Postigo, A., Fryer, C.~L., Page, K.~L., Gorosabel, J., Aloy, M.~A., Perley, D.~A., et al., 2011, Natur, 480, 72. doi:10.1038/nature10611

\bibitem[\protect\citeauthoryear{Thorne \& Zytkow}{1977}]{ThorneZytkow1977} {{{ Thorne K.~S., Zytkow A.~N., 1977, ApJ, 212, 832. doi:10.1086/155109 }}}

\bibitem[Turk et al.(2011)]{Turk2011} Turk, M.~J., Smith, B.~D., Oishi, J.~S., et al.\ 2011, \apjs, 192, 9 

\bibitem[Vick et al.(2021)]{Vicketal2021MNRAS} Vick, M., MacLeod, M., Lai, D., \& Loeb A.\ 2021, \mnras, 503, 5569. doi:10.1093/mnras/stab850

\bibitem[\protect\citeauthoryear{Yang et al.}{2021}]{Yangetal2020} Yang S., Sollerman J., Chen T.-W., Kool E.~C., Lunnan R., Schulze S., Strotjohann N., et al., 2021, A\&A, 646, A22. doi:10.1051/0004-6361/202039440

\bibitem[Zevin et al.(2021)]{Zevinetal2021} Zevin, M., Bavera, S.~S., Berry, C.~P.~L., et al.\ 2021, \apj, 910, 152. doi:10.3847/1538-4357/abe40e

\bibitem[Zheng \& Yu(2021)]{ZhengYu2021} Zheng, J.-H. \& Yu, Y.-W.\ 2021, Research in Astronomy and Astrophysics, 21, 200. doi:10.1088/1674-4527/21/8/200
\end{thebibliography}
\end{document}